\documentclass{ws-ijbc}
\usepackage{ws-rotating}     
\usepackage{graphicx}
\usepackage{epstopdf}
\begin{document}

\catchline{}{}{}{}{} 

\markboth{Katsanikas et al.}{Phase transport in a Caldera  with three index-1 saddles}

\title{Phase space transport in a symmetric Caldera potential with three index-1 saddles and no minima}

\author{Matthaios Katsanikas}
\address{Research Center for Astronomy and Applied Mathematics, Academy of Athens, Soranou Efesiou 4, Athens, GR-11527, Greece.\\
  School of Mathematics, University of Bristol, Fry Building, Woodland Road, Bristol, BS8 1UG, United Kingdom.\\ mkatsan@academyofathens.gr}

\author{Makrina Agaoglou}
\address{Instituto de Ciencias Matem{\'a}ticas, CSIC, C/Nicol{\'a}s Cabrera 15, Campus Cantoblanco, 28049 Madrid, Spain\\ makrina.agaoglou@icmat.es}

\author{Stephen Wiggins}
\address{School of Mathematics, University of Bristol, \\ Fry Building, Woodland Road, Bristol, BS8 1UG, United Kingdom.\\ s.wiggins@bristol.ac.uk}

\author{Ana M. Mancho}
\address{Instituto de Ciencias Matem{\'a}ticas, CSIC, C/Nicol{\'a}s Cabrera 15, Campus Cantoblanco, 28049 Madrid, Spain\\ a.m.mancho@icmat.es}

\maketitle

\begin{history}
\received{(to be inserted by publisher)}
\end{history}

\begin{abstract}
We apply the method of Lagrangian Descriptors (LDs) to a symmetric  Caldera-type potential energy surface which has  three index-1 saddles surrounding a relatively flat region that contains no minimum. Using this method we show the phase space transport mechanism that is responsible for the existence and non-existence of the phenomenon of dynamical matching for this form of Caldera potential energy surface.
\end{abstract}

\section{Introduction}
\label{intro}

In \cite{carpenter1985} Carpenter introduced a two dimensional (2D) potential energy surface (PES) which has been referred to as the ‘’Caldera’’ due to its topography resembling a collapsed volcano. In terms of its critical points the original Carpenter model was characterized by
one central minimum surrounded by four index-1 saddles that control the exit and entrance to the Caldera region or to  infinity  \cite{collins2014}. These types of potentials can be encountered in many organic chemical reactions (see, for example,  the introduction of \cite{collins2014,katsanikas2018}) and  PESs describing  four-armed barred galaxies \cite{athanassoula2009rings}. The dynamical properties associated with Caldera-type potentials have  been studied in many papers, such as \cite{katsanikas2018,katsanikas2019,katsanikas2020a,katsanikas2020b,geng2021bifurcations,geng2021influence,katsanikas2022nature}. 

In this paper we study the dynamics associated with a Caldera PES with a different critical point configuration.  It has only three index-1 saddles surrounding the Caldera region which contains a minimum. Two of the saddles are upper saddles, much like the original Caldera, and the remaining saddle is a lower saddle located in the center of the region.  Nevertheless, the lower saddle still  controls the exit of trajectories from the lower part of the Caldera region, as we will show.  This form of the  Caldera potential arises from the original Caldera PES when the critical points undergo a  pitchfork bifurcation, as described in \cite{geng2021bifurcations,geng2021influence}.

Dynamical matching is an important phenomenon in organic chemical reactions \cite{collins2014,katsanikas2018}. This phenomenon is encountered in the classical form of Caldera potential energy surface (i.e., with one central minimum and four index-1 saddles).  What dynamical matching means in this case is that  trajectories that are initiated in the region of the upper index-1  saddles, go straight across the caldera and exit through the region of the opposite lower saddle (see for example \cite{katsanikas2018}). In this paper we will study this phenomenon for the Caldera potential energy surface with three index-1 saddles \cite{geng2021bifurcations,geng2021influence}.

This paper is outlined as follows.
In section \ref{model} we describe the model that we use in this paper. In section \ref{phase} 
we analyze the phase space mechanism that is responsible for the existence and  non-existence of dynamical matching in the case of a Caldera potential energy surface with three index-1 saddles (see section \ref{phase}). We present our conclusions in section \ref{concl}. The method of Lagrangian descriptors plays an essential role in our analysis of phase space structure and the aspects of this method that we require are described in Appendix A.

\section{Model}
\label{model}

The model that we consider is the two dimensional (2D) symmetric Caldera potential energy surface (PES) introduced in \cite{collins2014}:

\begin{equation}
    \begin{aligned}
    V(x,y) &= c_1r^2+c_2y-c_3r^4\cos(4\theta)\\ &= c_1(x^2+y^2)+c_2y-c_3(x^4+y^4-6x^2y^2),
    \end{aligned}
\end{equation}

\noindent
where $(x,y)$, $(r,\theta)$ describe the position in Cartesian and polar coordinates, respectively, and $c_1,c_2,c_3$ are  parameters.

The corresponding 2 degree-of-freedom  (2 DoF) Hamiltonian is:

\begin{equation}
    H(x,y,p_x,p_y)=\frac{p_x^2}{2m}+\frac{p_y^2}{2m}+V(x,y),
\end{equation}

\noindent
where $p_x$ denotes the conjugate momentum corresponding to  $x$ and $p_y$ denotes the conjugate momentum corresponding to $y$. Moreover, we will assume that $m$ is $1$. Therefore, the equations of motion are:

\begin{equation}
    \begin{aligned}
    &\dot{x} = \frac{\partial H}{\partial p_x} = \frac{p_x}{m},\\
    &\dot{y} = \frac{\partial H}{\partial p_y} = \frac{p_y}{m},\\
    &\dot{p_x} = - \frac{\partial V}{\partial x}(x,y) = -(2c_1x-4c_3x^3+12c_3xy^2),\\
    &\dot{p_y} = - \frac{\partial V}{\partial y}(x,y) = -(2c_1y-4c_3y^3+12c_3x^2y+c_2).
    \end{aligned}
\end{equation}

In this paper we will study the case where $c_{1}=0.4,c_{2}=3,c_{3}=-0.3$.  For these parameter values there are three index-1 saddles. Two upper saddles, that we will refer to as upper left hand (LH) saddle and upper right hand (RH) saddle, with value of energy 2.321, and one lower, that we will refer to as the lower saddle, with value of energy -2.402.  The positions of the equilibrium points and the corresponding values of energy are given in table \ref{table:1}.

\begin{table}[htbp]	
	\tbl{Stationary points of the Caldera potential for $c_1=0.4,c_2=3$ and $c_3=-0.3$  ("RH" and "LH" are the abbreviations for right hand and left hand respectively).} {
	 \begin{tabular}{l  l  l  l}
    \hline
    Critical point & x & y & E \\
    \hline
    Lower  saddle & 0.000 & -1.194 & -2.402 \\
    Upper LH saddle  & -1.204 & 0.840  & 2.321 \\
    Upper RH saddle  & 1.204 &  0.840 & 2.321 \\
    \hline
    \end{tabular}
		\label{table:1} } 
\end{table}

The potential energy surface (PES) of our model  is depicted in Fig. \ref{fig:3DPES}:

\begin{figure}
	\centering
	
    \includegraphics[width=0.6\linewidth]{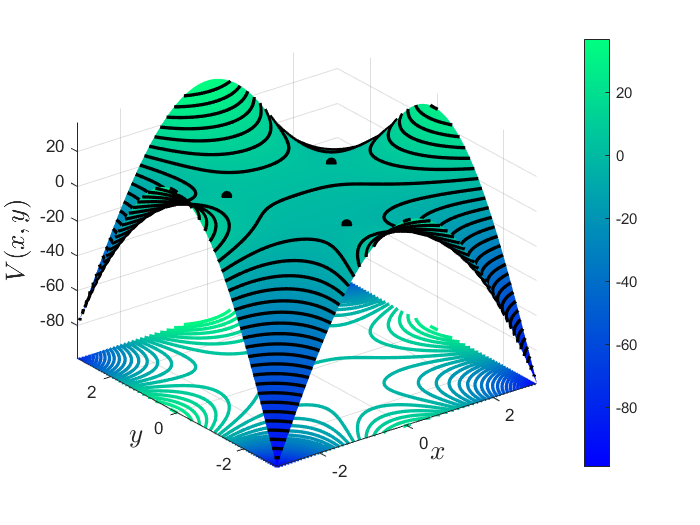}
	
	\caption{The 3D representation of the PES for $c_{1}=0.4,c_{2}=3,c_{3}=-0.3$}
	\label{fig:3DPES}
\end{figure}

The dynamical significance of the index-1 saddles is that for energies above that of the saddles, there exist unstable periodic orbits (UPOs) in phase space. This is a consequence of the Lyapunov subcenter theorem \cite{moser1958, kelley1967}. These UPOs, and their stable and unstable manifolds, are the essential phase space structures for understanding dynamical matching.

\section{Phase Space  Mechanism For the Existence and  Non-existence of Dynamical Matching}
\label{phase}

We begin by explaining what we mean by dynamical matching in this potential energy surface with three index-1 saddles. In the ‘’usual’’ Caldera potential energy surfaces  with two upper saddles and two lower saddles dynamical matching means that trajectories entering the Caldera region from the upper right saddle (resp. upper left saddle) move directly across the Caldera and exit by crossing the lower left saddle (resp. lower right saddle). For the Caldera potential energy surface with three index-1 saddles there is only a single lower index-1 saddle in the center of the lower regions. However, one can see from Fig. \ref{fig:3DPES} that there is an exit region to the left (left exit region, LER) of the saddle and an exit region to the right (right exit region, RER) of the saddle. In this case, dynamical matching corresponds to trajectories that enter the Caldera from the upper right saddle (resp. upper left saddle), move directly across the Caldera, and exit from the LER (resp. RER).

Both the existence and non-existence of dynamical matching in this model can be explained by heteroclinic connections between the unstable manifolds of the upper UPOs and the stable manifold of the UPO associated with the lower saddle (lower UPO) and the geometry of the unstable manifold of the lower UPO.

\begin{figure}
 \centering
 A)\includegraphics[scale=0.55]{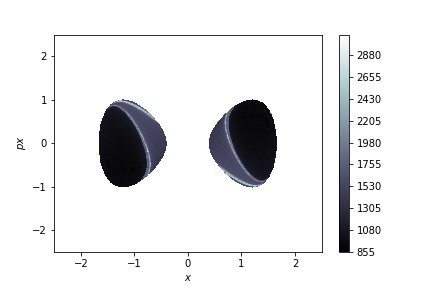}
 B)\includegraphics[scale=0.55]{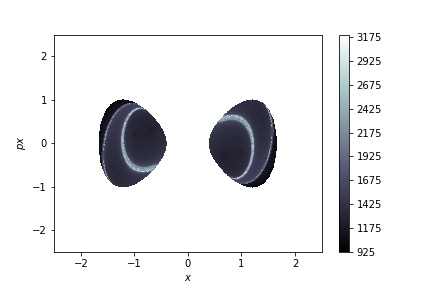}\\
 C)\includegraphics[scale=0.55]{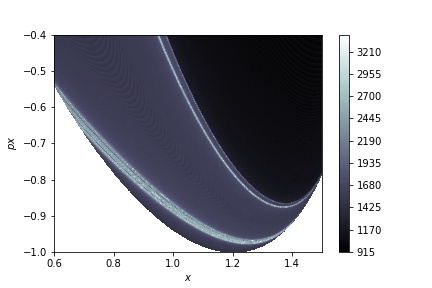}
 D)\includegraphics[scale=0.55]{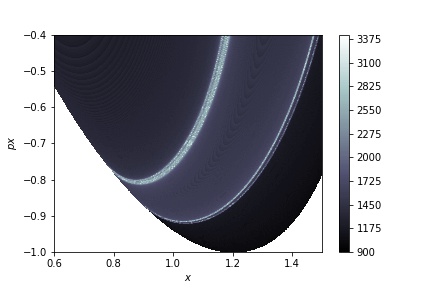}\\
\caption{Computation of variable-time LDs in the 2D slide $y= 0.84$ using $\tau = 8$ and p = 1/2 for value of energy $E=2.821$. A) Forward integration LDs;
B) Backward integration LDs; C) Forward integration for the interval $0.6<x<1.5$ and  $-1<p_x<-0.4$ . D) Backward integration for the interval $0.6<x<1.5$ and  $-1<p_x<-0.4$.}
\label{lds}
\end{figure}

We will study the geometry of these manifolds by using the method of Lagrangian descriptors (LDs), described in Appendix A. In order to study the manifolds near the upper saddles we apply the method of LDs in the 2D slice $y=0.84$ with $p_y>0$ (for a value of energy $E=2.821$) since this is the  $y$-coordinate of the upper index-1 saddles (see the panels A and B of  Fig. \ref{lds}).

This the geometric structures on this slice reveal the unstable manifolds of the unstable periodic orbits of the upper index-1 saddles. We will focus on the right hand side of this slice (see panels C and D of Fig. \ref{lds}) in order to study in detail the invariant manifolds that emanate from the unstable periodic orbits of the right upper index-1 saddle. In particular, we will describe the phase space mechanism for the case of the upper right index-1 saddle since for the case of the upper left saddle we  have  similar  mechanism as a result of  the symmetry of the potential. 

\begin{figure}
 \centering
 \includegraphics[scale=0.65]{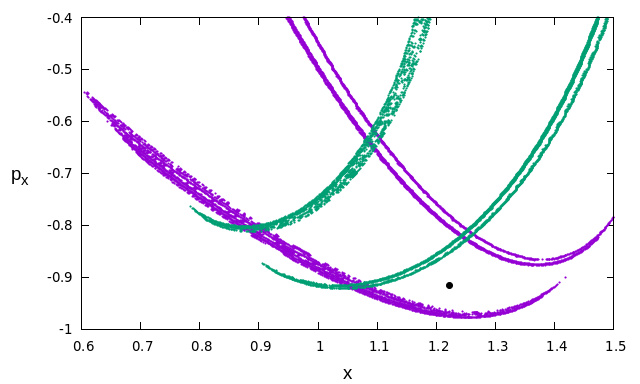}\\
\caption{The stable (violet) and unstable (green) invariant manifolds extracted (in the 2D slice $y= 0.84$ with $p_y>0$) from the gradient of the Lagrangian descriptors for the interval $0.6<x<1.5$ and  $-1<p_x<-0.4$. We depict using black color the initial condition of a  trajectory.}
\label{lds-1}
\end{figure}

\begin{figure}
 \centering
 \includegraphics[scale=0.65]{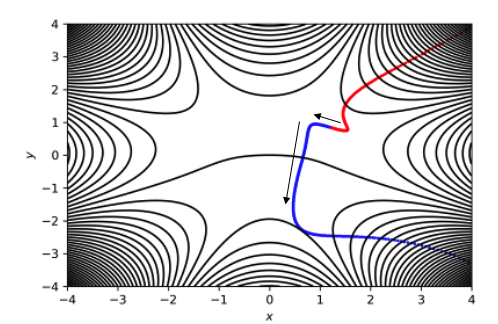}\\
\caption{The contours of the Caldera potential and a trajectory in the configuration space forward (with blue color) and backward (with red color) in time (the initial condition of this trajectory is depicted by a black point in Fig. \ref{lds-1}). The arrows denote the direction of the trajectory forward in time. }
\label{orb}
\end{figure}

Next we extract the invariant manifolds of the panels C and D of Fig. \ref{lds} using the gradient of LDs (see details in \cite{katsanikas2020b}). We can see the result in Fig. \ref{lds-1}. In this figure, we see many lobes between unstable and stable invariant manifolds. This is the unstable manifold of the UPO associated with the upper right saddle and the stable manifold associated the UPO of the lower saddle.  We can verify this as follows.  In the right corner of this figure we depict a position of a trajectory inside a lobe. If we integrate this trajectory backwards and forwards, the result will be the exit of this trajectory from the caldera through the region of the upper right index-1 saddle and the region of the lower right exit (see Fig. \ref{orb}), respectively. This means that this lobe corresponds to a lobe between intersections of the unstable invariant manifold of the unstable periodic orbits of the UPO of the upper right index-1 saddle with the stable invariant manifold  of the UPO family of the  lower index-1 saddle (that is located at the center). The trajectories that are initially at the region of the right upper saddle follow  the unstable invariant manifolds of the UPO associated with the  right upper saddle until they are trapped in the lobes between the heteroclinic intersections of the unstable invariant manifolds of the right upper index-1 saddle with the stable invariant manifolds of the UPO of the lower index-1 saddle. Then the trajectories follow the stable invariant manifolds to the area of these unstable periodic orbits (the central area of the Caldera). This is the reason that the trajectory that corresponds to the black point of Fig. \ref{lds-1}, moves towards  the central area forward in time (as indicated from the arrows in Fig. \ref{orb}). This explains the first part of the mechanism, the transport of the trajectories from the region of the upper index-1 saddle to the  central area of the caldera.

\begin{figure}
 \centering
 A)\includegraphics[scale=0.55]{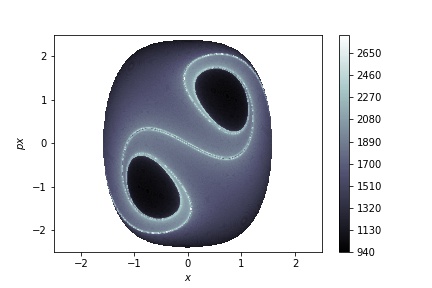}
 B)\includegraphics[scale=0.55]{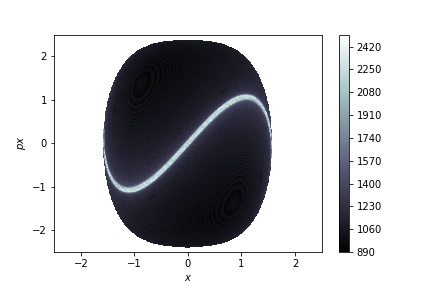}\\
\caption{Computation of variable-time LDs in the 2D slice $y= 0$ using $\tau = 5$ and p = 1/2 for value of energy $E=2.821$. A) Forward integration LDs;
B) Backward integration LDs;} 
\label{lds2}
\end{figure}

\begin{figure}
 \centering
 \includegraphics[scale=0.75]{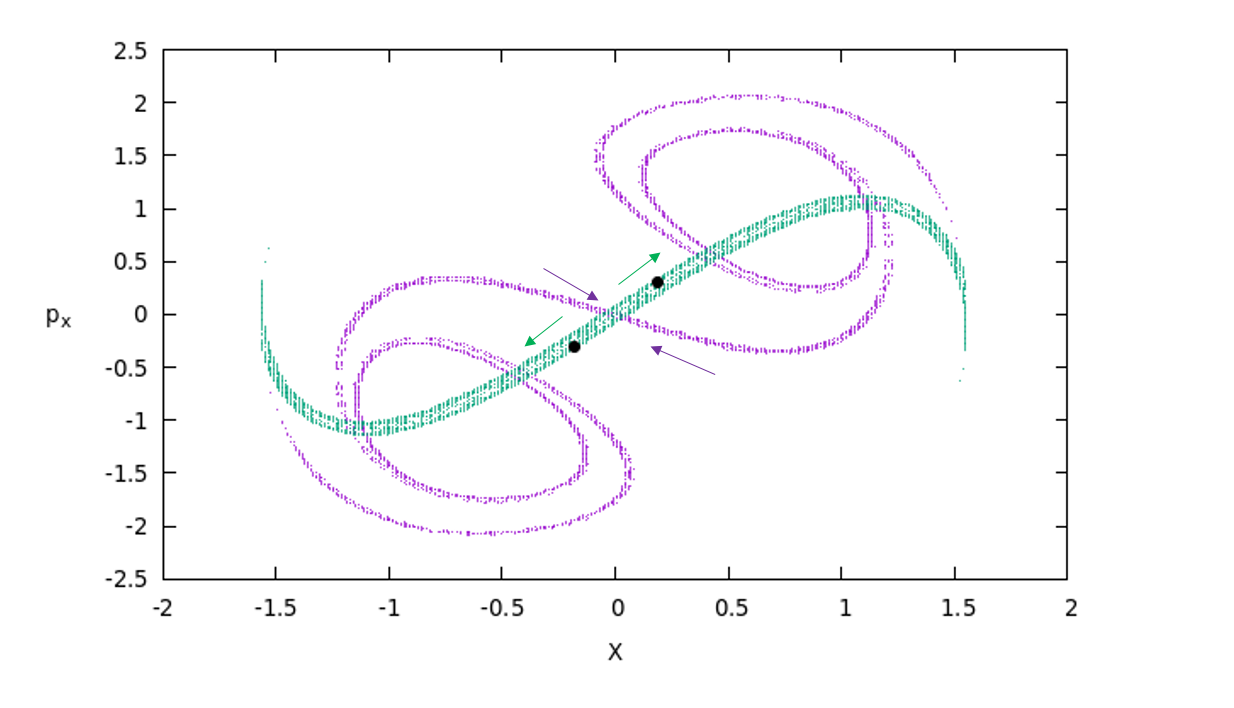}\\
\caption{The stable (violet) and unstable (green) invariant manifolds extracted (in the 2D slice $y= 0$ with $p_y>0$) from the gradient of the Lagrangian descriptors for the interval $-2.5<x<2.5$ and  $-2.5<p_x<2.5$. We depict using black color the initial conditions of two  trajectories on the unstable manifold. The green and violet arrows present the directions of the unstable and stable invariant manifolds respectively that are emanated from the unstable periodic orbit of the lower index-1, that is located at the center.}
\label{lds-2}
\end{figure}

\begin{figure}
 \centering
 A)\includegraphics[scale=0.65]{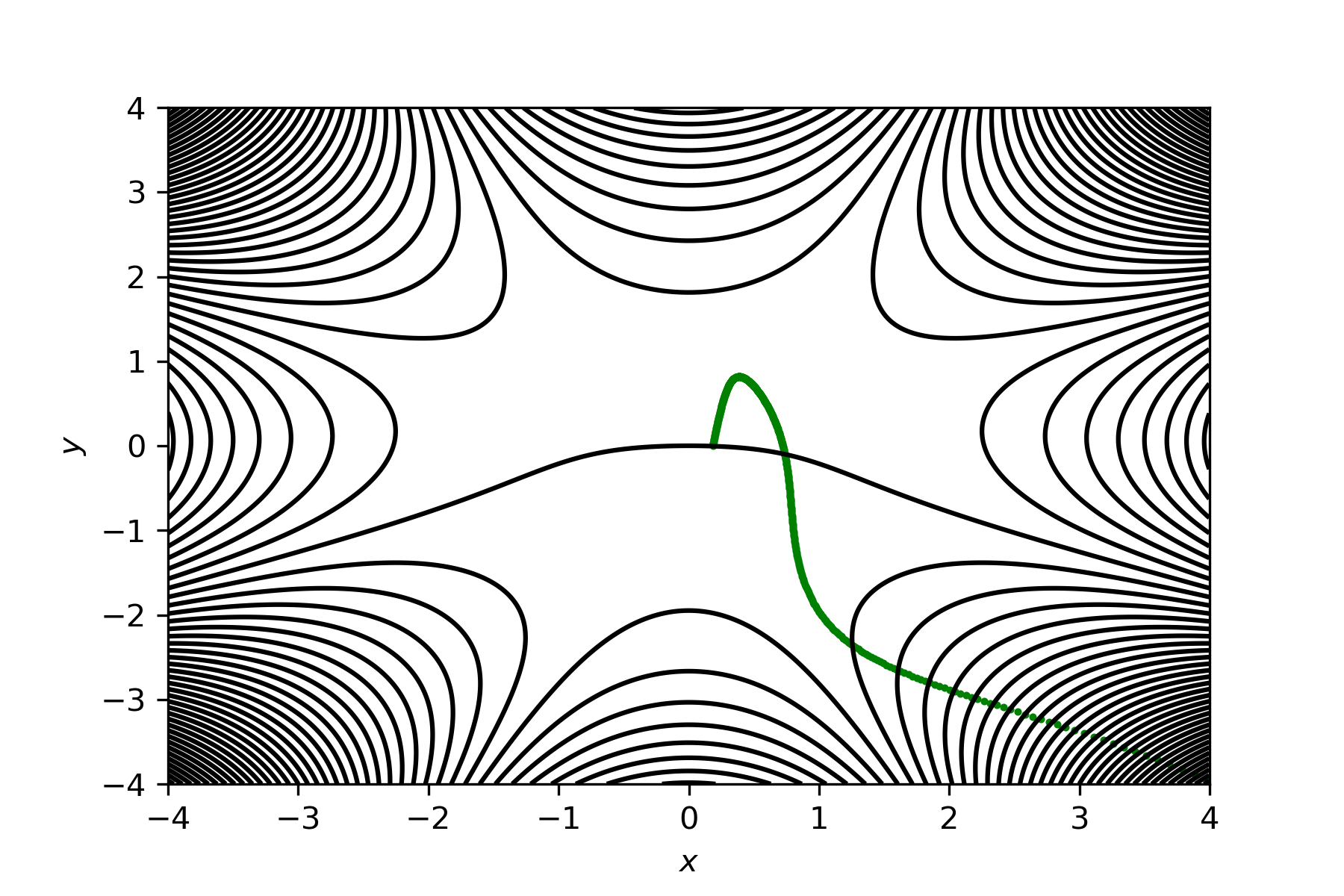}
  B)\includegraphics[scale=0.65]{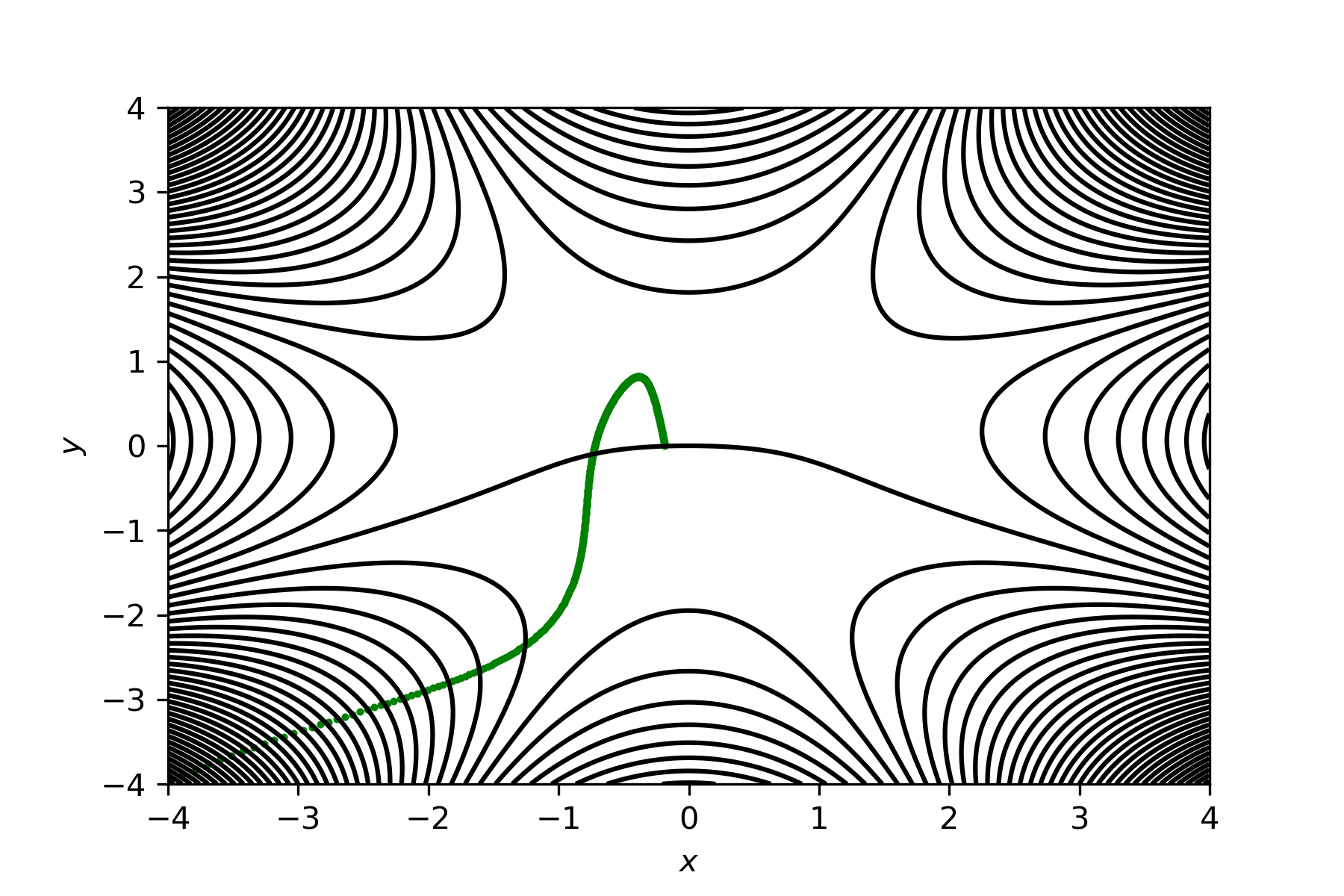}\\
\caption{The contours of the caldera potential and two trajectories in the configuration space forward in time (with green  color). The initial conditions of these trajectories correspond to the  two black points in the  right side from the center in Fig. \ref{lds-2} (for panel A) and in the  left  side from the center in Fig. \ref{lds-2} (for panel B). }
\label{orb2}
\end{figure}

The second part of the mechanism explains the transport of the trajectories from the area of the lower index-1 saddle (central area of the caldera)  to one of the lower  exits from the Caldera (for example the lower right exit from the Caldera for the trajectory of Fig. \ref{orb}). In order to see this we use the method of LDs in the slice $y=0$ with $p_y>0$ to detect the invariant manifolds emanating from the central area of the caldera and, in particular, the invariant manifolds of the lower index-1 saddle. In Fig. \ref{lds2}, we depict, using LDs, the unstable and stable invariant manifolds of the central area. The next step is to extract the invariant manifolds of the panels A and B of Fig. \ref{lds2} using the gradient of LDs (see details in \cite{katsanikas2020b}). We can see the result in Fig. \ref{lds-2}. In this Figure, we see the unstable and stable manifolds start from the point of the periodic orbit of the lower index-1 saddle (that is located at the center). The trajectories that approach the periodic orbit are on the stable manifolds (the motion of these trajectories is indicated by the violet arrows in Fig. \ref{lds-2}). The trajectories that move away from the periodic orbit are on the unstable manifolds (the motion of these trajectories is indicated by the green arrows in Fig. \ref{lds-2}. This means that the trajectories in the first part of the mechanism 
approach the neighborhood of the periodic orbit of the lower index-1 saddle (that is located in the central area of the caldera) will leave from this neighborhood following the unstable manifolds in one of the two directions that are indicated by the green arrows in Fig. \ref{lds-2}. In each of these two directions, we take one initial condition (see the black points in Fig. \ref{lds-2}) and we integrate it forward in time. In this way, we want to check the fate of the trajectories that will follow the one or the other direction of the unstable manifolds. When we integrate these initial conditions, we observe in the panel A of Fig. \ref{orb2} that the trajectory that corresponds to the black point on the right side from the center (where the periodic orbit of the lower index-1 saddle is located) exits through the region of the lower right exit. On the contrary, when we integrate the other initial condition (the black point on the left side from the center of Fig. \ref{lds-2}) the corresponding trajectory exits through the region of the lower left exit (see panel B of Fig. \ref{orb2}). We see that the two directions of the unstable manifolds that are emanated from the unstable periodic orbit of the lower index-1 saddle guide the trajectories directly to different exit regions. This means that the trajectories that are in the neighborhood of the unstable periodic orbits of the lower index-1 saddles in the central area of the caldera leave this area following the unstable invariant manifolds of these periodic orbits. If they follow the one direction of the unstable invariant manifolds exit through the one lower exit region and if they follow the other direction, they exit through the other lower exit region. This indicates the important role of the periodic orbits of the lower index-1 saddle that control through the direction of the unstable invariant manifolds the exit of the trajectories that come from the region of the upper index-1 saddles to the central area of the caldera.

\section{Conclusions}
\label{concl}

In this paper, we investigated the phase space mechanism of transport that is
responsible for the creating and breaking  the phenomenon of dynamical
matching (that was observed in \cite{geng2021bifurcations,geng2021influence})
for the form of a Caldera potential energy surface with only three index-1
saddles and no minimum. The first part of the mechanism is based on
heteroclinic intersections of the unstable invariant manifolds of the
unstable periodic orbits of the upper index-1 saddles and the stable invariant
manifolds of the unstable periodic orbits of the lower index-1 saddle. This
explains the transport from the region of the upper index-1 saddles to the
region of the lower index-1 saddle.  The second part of the mechanism is based
on the direction of the unstable invariant manifolds of the unstable periodic
orbits of the lower index-1 saddle. The trajectories leave the central area of
the caldera following the unstable invariant manifolds of the unstable
periodic orbits of the lower index-1 saddle. This happens in two directions.
One direction guides the trajectories to the region of the lower left exit
region and the other to the region of the lower right exit region. The
unstable periodic orbits of the lower index-1 saddle control the transport
from the central region of the caldera to the lower exit regions.

\nonumsection{Acknowledgments}The authors acknowledge the financial support provided by the EPSRC Grant No. EP/P021123/1 and MA acknowledges support from the grant CEX2019-000904-S and IJC2019-040168-I funded by: MCIN/AEI/ 10.13039/501100011033.

\bibliographystyle{ws-ijbc}
\bibliography{caldera2c}

\begin{thebibliography}{24}
\newcommand{\enquote}[1]{``#1''}
\providecommand{\natexlab}[1]{#1}
\providecommand{\url}[1]{\texttt{#1}}
\providecommand{\urlprefix}{URL }
\expandafter\ifx\csname urlstyle\endcsname\relax
  \providecommand{\doi}[1]{doi:\discretionary{}{}{}#1}\else
  \providecommand{\doi}{doi:\discretionary{}{}{}\begingroup
  \urlstyle{rm}\Url}\fi

\bibitem[{Agaoglou \emph{et~al.}(2019)Agaoglou, Aguilar-Sanjuan,
  Garc{\'i}a-Garrido, Garc{\'i}a-Meseguer, Gonz{\'a}lez-Montoya, Katsanikas,
  Krajňák, Naik \& Wiggins}]{Agaoglou2019}
Agaoglou, M., Aguilar-Sanjuan, B., Garc{\'i}a-Garrido, V.~J.,
  Garc{\'i}a-Meseguer, R., Gonz{\'a}lez-Montoya, F., Katsanikas, M., Krajňák,
  V., Naik, S. \& Wiggins, S. [2019] \emph{Chemical Reactions: A Journey into
  Phase Space} (Bristol, UK, Zenodo), \doi{10.5281/zenodo.3568210}.

\bibitem[{Agaoglou \emph{et~al.}(2020)Agaoglou, Aguilar-Sanjuan,
  Garc{\'i}a-Garrido, Gonz{\'a}lez-Montoya, Katsanikas, Krajňák, Naik \&
  Wiggins}]{ldbook2020}
Agaoglou, M., Aguilar-Sanjuan, B., Garc{\'i}a-Garrido, V.~J.,
  Gonz{\'a}lez-Montoya, F., Katsanikas, M., Krajňák, V., Naik, S. \& Wiggins,
  S. [2020] \emph{Lagrangian Descriptors: Discovery and Quantification of Phase
  Space Structure and Transport} (zenodo: 10.5281/zenodo.3958985),
  \doi{10.5281/zenodo.3958985}.

\bibitem[{Athanassoula \emph{et~al.}(2009)Athanassoula, Romero-G{\'o}mez, Bosma
  \& Masdemont}]{athanassoula2009rings}
Athanassoula, E., Romero-G{\'o}mez, M., Bosma, A. \& Masdemont, J. [2009]
  \enquote{Rings and spirals in barred galaxies--ii. ring and spiral
  morphology,} \emph{Monthly Notices of the Royal Astronomical Society}
  \textbf{400},  1706--1720.

\bibitem[{Balibrea-Iniesta \emph{et~al.}(2016)Balibrea-Iniesta, Lopesino,
  Wiggins \& Mancho}]{balibrea2016}
Balibrea-Iniesta, F., Lopesino, C., Wiggins, S. \& Mancho, A.~M. [2016]
  \enquote{Lagrangian descriptors for stochastic differential equations: A tool
  for revealing the phase portrait of stochastic dynamical systems,}
  \emph{International Journal of Bifurcation and Chaos} \textbf{26},  1630036,
  \doi{10.1142/S0218127416300366}.

\bibitem[{Balibrea-Iniesta \emph{et~al.}(2019)Balibrea-Iniesta, Xie,
  Garc\'{i}a-Garrido, Bertino, Mancho \& Wiggins}]{balibrea2019}
Balibrea-Iniesta, F., Xie, J., Garc\'{i}a-Garrido, V.~J., Bertino, L., Mancho,
  A.~M. \& Wiggins, S. [2019] \enquote{Lagrangian transport across the upper
  arctic waters in the canada basin,} \emph{Quarterly Journal of the Royal
  Meteorological Society} \textbf{145},  76--91, \doi{10.1002/qj.3404}.

\bibitem[{Carpenter(1985)}]{carpenter1985}
Carpenter, B.~K. [1985] \enquote{Trajectories through an intermediate at a
  fourfold branch point. implications for the stereochemistry of biradical
  reactions,} \emph{Journal of the American Chemical Society} \textbf{107},
  5730--5732, \doi{10.1021/ja00306a021}.

\bibitem[{Collins \emph{et~al.}(2014)Collins, Kramer, Carpenter, Ezra \&
  Wiggins}]{collins2014}
Collins, P., Kramer, Z., Carpenter, B., Ezra, G. \& Wiggins, S. [2014]
  \enquote{Nonstatistical dynamics on the caldera,} \emph{Journal of Chemical
  Physics} \textbf{141}.

\bibitem[{Crossley \emph{et~al.}(2021)Crossley, Agaoglou, Katsanikas \&
  Wiggins}]{crossley2021poincare}
Crossley, R., Agaoglou, M., Katsanikas, M. \& Wiggins, S. [2021] \enquote{From
  poincar{\'e} maps to lagrangian descriptors: The case of the valley ridge
  inflection point potential,} \emph{Regular and Chaotic Dynamics} \textbf{26},
   147--164.

\bibitem[{Garc{\'i}a-Garrido \emph{et~al.}(2018)Garc{\'i}a-Garrido, Curbelo,
  Mancho, Wiggins \& Mechoso}]{gg2018}
Garc{\'i}a-Garrido, V.~J., Curbelo, J., Mancho, A.~M., Wiggins, S. \& Mechoso,
  C.~R. [2018] \enquote{The application of lagrangian descriptors to 3d vector
  fields,} \emph{Regul Chaotic Dyn} \textbf{23},  551--568,
  \doi{10.1134/S1560354718050052}.

\bibitem[{Geng \emph{et~al.}(2021{\natexlab{a}})Geng, Katsanikas, Agaoglou \&
  Wiggins}]{geng2021bifurcations}
Geng, Y., Katsanikas, M., Agaoglou, M. \& Wiggins, S. [2021{\natexlab{a}}]
  \enquote{The bifurcations of the critical points and the role of the depth in
  a symmetric caldera potential energy surface,} \emph{Int. J. Bifurcation
  Chaos} \textbf{31}.

\bibitem[{Geng \emph{et~al.}(2021{\natexlab{b}})Geng, Katsanikas, Agaoglou \&
  Wiggins}]{geng2021influence}
Geng, Y., Katsanikas, M., Agaoglou, M. \& Wiggins, S. [2021{\natexlab{b}}]
  \enquote{The influence of a pitchfork bifurcation of the critical points of a
  symmetric caldera potential energy surface on dynamical matching,}
  \emph{Chemical Physics Letters} \textbf{768},  138397.

\bibitem[{Katsanikas \emph{et~al.}(2020{\natexlab{a}})Katsanikas,
  Garc{\'\i}a-Garrido, Agaoglou \& Wiggins}]{katsanikas2020phase}
Katsanikas, M., Garc{\'\i}a-Garrido, V.~J., Agaoglou, M. \& Wiggins, S.
  [2020{\natexlab{a}}] \enquote{Phase space analysis of the dynamics on a
  potential energy surface with an entrance channel and two potential wells,}
  \emph{Physical Review E} \textbf{102},  012215.

\bibitem[{Katsanikas \emph{et~al.}(2020{\natexlab{b}})Katsanikas,
  Garc\'{i}a-Garrido \& Wiggins}]{katsanikas2020b}
Katsanikas, M., Garc\'{i}a-Garrido, V.~J. \& Wiggins, S. [2020{\natexlab{b}}]
  \enquote{Detection of dynamical matching in a caldera hamiltonian system
  using lagrangian descriptors,} \emph{Int. J. Bifurcation Chaos} \textbf{30},
  2030026.

\bibitem[{Katsanikas \emph{et~al.}(2020{\natexlab{c}})Katsanikas,
  Garc\'{i}a-Garrido \& Wiggins}]{katsanikas2020a}
Katsanikas, M., Garc\'{i}a-Garrido, V.~J. \& Wiggins, S. [2020{\natexlab{c}}]
  \enquote{The dynamical matching mechanism in phase space for caldera-type
  potential energy surfaces,} \emph{Chemical Physics Letters} \textbf{743},
  137199, \doi{https://doi.org/10.1016/j.cplett.2020.137199}.

\bibitem[{Katsanikas \& Wiggins(2018)}]{katsanikas2018}
Katsanikas, M. \& Wiggins, S. [2018] \enquote{Phase space structure and
  transport in a caldera potential energy surface,} \emph{International Journal
  of Bifurcation and Chaos} \textbf{28},  1830042.

\bibitem[{Katsanikas \& Wiggins(2019)}]{katsanikas2019}
Katsanikas, M. \& Wiggins, S. [2019] \enquote{Phase space analysis of the
  nonexistence of dynamical matching in a stretched caldera potential energy
  surface,} \emph{International Journal of Bifurcation and Chaos} \textbf{29},
  1950057.

\bibitem[{Katsanikas \& Wiggins(2022)}]{katsanikas2022nature}
Katsanikas, M. \& Wiggins, S. [2022] \enquote{The nature of reactive and
  non-reactive trajectories for a three dimensional caldera potential energy
  surface,} \emph{Physica D: Nonlinear Phenomena} ,  133293.

\bibitem[{Kelley(1967)}]{kelley1967}
Kelley, A. [1967] \enquote{On the {L}iapounov subcenter manifold,}
  \emph{Journal of mathematical analysis and applications} \textbf{18},
  472--478.

\bibitem[{Lopesino \emph{et~al.}(2017)Lopesino, Balibrea-Iniesta,
  Garc\'ia-Garrido, Wiggins \& Mancho}]{lopesino2017}
Lopesino, C., Balibrea-Iniesta, F., Garc\'ia-Garrido, V.~J., Wiggins, S. \&
  Mancho, A.~M. [2017] \enquote{A theoretical {F}ramework for {L}agrangian
  {D}escriptors,} \emph{Int J Bifurc Chaos} \textbf{27},  1730001,
  \doi{10.1142/S0218127417300014}.

\bibitem[{Lopesino \emph{et~al.}(2015)Lopesino, Balibrea-Iniesta, Wiggins \&
  Mancho}]{lopesino2015}
Lopesino, C., Balibrea-Iniesta, F., Wiggins, S. \& Mancho, A.~M. [2015]
  \enquote{Lagrangian descriptors for two dimensional, area preserving,
  autonomous and nonautonomous maps,} \emph{Communications in Nonlinear Science
  and Numerical Simulation} \textbf{27},  40--51,
  \doi{https://doi.org/10.1016/j.cnsns.2015.02.022}.

\bibitem[{Madrid \& Mancho(2009)}]{madrid2009}
Madrid, J. A.~J. \& Mancho, A.~M. [2009] \enquote{{Distinguished trajectories
  in time dependent vector fields},} \emph{Chaos} \textbf{19},  013111,
  \doi{10.1063/1.3056050}.

\bibitem[{Mancho \emph{et~al.}(2013)Mancho, Wiggins, Curbelo \&
  Mendoza}]{mancho2013lagrangian}
Mancho, A.~M., Wiggins, S., Curbelo, J. \& Mendoza, C. [2013]
  \enquote{Lagrangian descriptors: A method for revealing phase space
  structures of general time dependent dynamical systems,} \emph{Commun.
  Nonlinear Sci. Numer. Simul.} \textbf{18},  3530--3557,
  \doi{https://doi.org/10.1016/j.cnsns.2013.05.002}.

\bibitem[{Mendoza \& Mancho(2010)}]{Mancho1}
Mendoza, C. \& Mancho, A.~M. [2010] \enquote{The hidden geometry of ocean
  flows,} \emph{Phys. Rev. Lett.} \textbf{105},  038501,
  \doi{https://journals.aps.org/prl/abstract/10.1103/PhysRevLett.105.038501}.

\bibitem[{Moser(1958)}]{moser1958}
Moser, J. [1958] \enquote{On the generalization of a theorem of {A}.
  {L}iapounoff,} \emph{Communications on Pure and Applied Mathematics}
  \textbf{11},  257--271.

\end{thebibliography}


\begin{thebibliography}{12}
\newcommand{\enquote}[1]{``#1''}
\providecommand{\natexlab}[1]{#1}
\providecommand{\url}[1]{\texttt{#1}}
\providecommand{\urlprefix}{URL }
\expandafter\ifx\csname urlstyle\endcsname\relax
  \providecommand{\doi}[1]{doi:\discretionary{}{}{}#1}\else
  \providecommand{\doi}{doi:\discretionary{}{}{}\begingroup
  \urlstyle{rm}\Url}\fi

\bibitem[{Athanassoula \emph{et~al.}(2009)Athanassoula, Romero-G{\'o}mez, Bosma
  \& Masdemont}]{athanassoula2009rings}
Athanassoula, E., Romero-G{\'o}mez, M., Bosma, A. \& Masdemont, J. [2009]
  \enquote{Rings and spirals in barred galaxies--ii. ring and spiral
  morphology,} \emph{Monthly Notices of the Royal Astronomical Society}
  \textbf{400},  1706--1720.

\bibitem[{Carpenter(1985)}]{carpenter1985}
Carpenter, B.~K. [1985] \enquote{Trajectories through an intermediate at a
  fourfold branch point. implications for the stereochemistry of biradical
  reactions,} \emph{Journal of the American Chemical Society} \textbf{107},
  5730--5732, \doi{10.1021/ja00306a021}.

\bibitem[{Collins \emph{et~al.}(2014)Collins, Kramer, Carpenter, Ezra \&
  Wiggins}]{collins2014}
Collins, P., Kramer, Z., Carpenter, B., Ezra, G. \& Wiggins, S. [2014]
  \enquote{Nonstatistical dynamics on the caldera,} \emph{Journal of Chemical
  Physics} \textbf{141}.

\bibitem[{Geng \emph{et~al.}(2021{\natexlab{a}})Geng, Katsanikas, Agaoglou \&
  Wiggins}]{geng2021bifurcations}
Geng, Y., Katsanikas, M., Agaoglou, M. \& Wiggins, S. [2021{\natexlab{a}}]
  \enquote{The bifurcations of the critical points and the role of the depth in
  a symmetric caldera potential energy surface,} \emph{Int. J. Bifurcation
  Chaos} \textbf{31}.

\bibitem[{Geng \emph{et~al.}(2021{\natexlab{b}})Geng, Katsanikas, Agaoglou \&
  Wiggins}]{geng2021influence}
Geng, Y., Katsanikas, M., Agaoglou, M. \& Wiggins, S. [2021{\natexlab{b}}]
  \enquote{The influence of a pitchfork bifurcation of the critical points of a
  symmetric caldera potential energy surface on dynamical matching,}
  \emph{Chemical Physics Letters} \textbf{768},  138397.

\bibitem[{Katsanikas \emph{et~al.}(2020{\natexlab{a}})Katsanikas,
  Garc\'{i}a-Garrido \& Wiggins}]{katsanikas2020b}
Katsanikas, M., Garc\'{i}a-Garrido, V.~J. \& Wiggins, S. [2020{\natexlab{a}}]
  \enquote{Detection of dynamical matching in a caldera hamiltonian system
  using lagrangian descriptors,} \emph{Int. J. Bifurcation Chaos} \textbf{30},
  2030026.

\bibitem[{Katsanikas \emph{et~al.}(2020{\natexlab{b}})Katsanikas,
  Garc\'{i}a-Garrido \& Wiggins}]{katsanikas2020a}
Katsanikas, M., Garc\'{i}a-Garrido, V.~J. \& Wiggins, S. [2020{\natexlab{b}}]
  \enquote{The dynamical matching mechanism in phase space for caldera-type
  potential energy surfaces,} \emph{Chemical Physics Letters} \textbf{743},
  137199, \doi{https://doi.org/10.1016/j.cplett.2020.137199}.

\bibitem[{Katsanikas \& Wiggins(2018)}]{katsanikas2018}
Katsanikas, M. \& Wiggins, S. [2018] \enquote{Phase space structure and
  transport in a caldera potential energy surface,} \emph{International Journal
  of Bifurcation and Chaos} \textbf{28},  1830042.

\bibitem[{Katsanikas \& Wiggins(2019)}]{katsanikas2019}
Katsanikas, M. \& Wiggins, S. [2019] \enquote{Phase space analysis of the
  nonexistence of dynamical matching in a stretched caldera potential energy
  surface,} \emph{International Journal of Bifurcation and Chaos} \textbf{29},
  1950057.

\bibitem[{Katsanikas \& Wiggins(2022)}]{katsanikas2022nature}
Katsanikas, M. \& Wiggins, S. [2022] \enquote{The nature of reactive and
  non-reactive trajectories for a three dimensional caldera potential energy
  surface,} \emph{Physica D: Nonlinear Phenomena} ,  133293.

\bibitem[{Kelley(1967)}]{kelley1967}
Kelley, A. [1967] \enquote{On the {L}iapounov subcenter manifold,}
  \emph{Journal of mathematical analysis and applications} \textbf{18},
  472--478.

\bibitem[{Moser(1958)}]{moser1958}
Moser, J. [1958] \enquote{On the generalization of a theorem of {A}.
  {L}iapounoff,} \emph{Communications on Pure and Applied Mathematics}
  \textbf{11},  257--271.

\end{thebibliography}

\section{appendix}

The Lagrangian Descriptors is a diagnostic tool that can reveal the phase space structures. The first time that this technique was used was in the paper \cite{madrid2009}, were was aiming to study transport
and mixing in geophysical flows. In the last years this technique has been broadly used not only in fluid mechanics \cite{lopesino2015,balibrea2016,Mancho1,mancho2013lagrangian,gg2018,balibrea2019,lopesino2017} but also in the area of chemical reaction dynamics \cite{Agaoglou2019,ldbook2020,Agaoglou2019,katsanikas2020phase,crossley2021poincare}. The LDs method works in the following way: In order to reveal the phase space structures in a given slice using the method of LDs the steps that we need to follow are very simple. First we choose the slice and we define in this slice a grid of initial conditions. Then we integrate this initial conditions forward and backward in time for a given integration time $\tau$ and while we are integrating we  accumulate along these trajectories a positive quantity defined from the vector field that determines the dynamical system that we are studying. If you integrate trajectories forwards in time, the LD function is going to detect the stable manifolds while the backwards in time integration will detect  the unstable manifolds. The scalar output obtained from the method will highlight the location of the invariant stable and unstable manifolds intersecting this slice, which are detected at points where the values of LDs display an abrupt change. 

There are several definitions for the M function. In this work we are using the p-norm definition of the method that relies on variable time integration, where $p\in (0,1]$ (the reader can find more information about the application of variable time LDs to Caldera potentials in \cite{katsanikas2020b}). We have fixed the value of p to be $p=1/2$. This definition of the M function is preferable here due to the nature of the Caldera's potential energy surface (PES), that is an open potential. That can lead to an increasingly fast pace escape of the trajectories.

Let's consider the following dynamical system with time dependence:
\begin{equation}
\dfrac{d\mathbf{x}}{dt} = \mathbf{v}(\mathbf{x},t) \;,\quad \mathbf{x} \in \mathbb{R}^{n} \;,\; t \in \mathbb{R} \;
\label{eq:gtp_dynSys}
\end{equation}
where $\mathbf{v}(\mathbf{x},t) \in C^{r} (r \geq 1)$ in $\mathbf{x}$ and it is continuous in time. Given an initial condition $x_0$ at time $t_0$, take a fixed integration time $\tau>0$ and $p \in (0, 1]$. The method of LDs in the p-norm definition is as follows:

\begin{equation}
M_p(\mathbf{x}_{0},t_0,\tau) = \sum_{k=1}^{n} \bigg[ \int^{t_0+\tau}_{t_0-\tau}  |v_{k}(\mathbf{x}(t;\mathbf{x}_0),t)|^p \; dt \bigg] = M_p^{(b)}(\mathbf{x}_{0},t_0,\tau)+ M_p^{(f)}(\mathbf{x}_{0},t_0,\tau) \;, 
\label{eq:Mp_function}
\end{equation}

where $M_p^{(b)}$ and $M_p^{(f)}$ its backward and forward integration parts:
\begin{equation}
\begin{split}
M_p^{(b)}(\mathbf{x}_{0},t_0,\tau) &= \sum_{k=1}^{n} \bigg[ \int^{t_0}_{t_0-\tau}  |v_{k}(\mathbf{x}(t;\mathbf{x}_0),t)|^p \; dt \bigg] \;, \\[.2cm]
M_p^{(f)}(\mathbf{x}_{0},t_0,\tau) &= \sum_{k=1}^{n} \bigg[ \int^{t_0+\tau}_{t_0} |v_{k}(\mathbf{x}(t;\mathbf{x}_0),t)|^p \; dt \bigg] \;,
\end{split}
\end{equation}

The formulation of the p-norm definition that we apply to this model is the following:

\begin{equation}
    M_p(\mathbf{x}_{0},t_0,\tau) = \sum_{k=1}^{n} \bigg[ \int^{t_0+\tau_{x_{0}}^{+}}_{t_0-\tau_{x_{0}}^{-}}  |v_{k}(\mathbf{x}(t;\mathbf{x}_0),t)|^p \; dt \bigg]
\end{equation}

where 
$$\tau_{x_{0}}^{\pm}=min\{\tau_{0},|t^{\pm}|_{|x(t^{\pm};x_{0}\notin R)}\}$$

Note that the integration time $\tau_{0}$ is fixed and that $t^{+}$ is the time that the trajectory exits the interaction region $R$ in forward time whereas $t^{-}$ is for backward time.

\end{document}